\documentclass[12pt]{article}
\usepackage{graphicx}  % standard LaTeX graphics tool
\usepackage{graphics}  % standard LaTeX graphics tool
\usepackage{subfig}
\usepackage{amssymb}
\usepackage{amsmath}
\usepackage{epsfig}
\begin{document}

\title{On Ultrafast Spin Dynamics: Spin Dependent Fast Response of Hot Electrons, of Band--Structure}

\author{M.Avignon, \\Institut Neel, CNRS and Universite J.Fourier, BP166, 38042 Grenoble\\
K.H.Bennemann,\\ Institute for Theoretical Physics, Freie Universit\"{a}t Berlin,\\ Arnimallee 14, 14195 Berlin}
\maketitle

\abstract{We analyze the fast spin dynamics of hot electrons. For times of the order of the spin dependent lifetimes of the hot electrons spin dependent response occurs. In particular, different energy shifts for majority and minority electrons are expected in general. Thus, for example in case of (laser) excited ferromagnetic metals majority and minority electrons may respond differently in time during closing the exchange splitting. Spin flip transitions of the hot electrons due to electron interactions cause quasi hybridization of the spin split states. This is also the case in itinerant ferromagnetic metals due to hopping between sites having magnetic moments pointing in direction of the magnetization (+) and opposite direction (-) and with energy levels $\varepsilon^+_{i\sigma}$ and $\varepsilon^-_{i\sigma}$. For energetic reasons the molecular field acts asymmetrically on the spins of the electrons and on spin flip transitions and thus causes different lifetimes of minority and majority electrons and spin dependent electron energy shifts.  Quite general minority hot electrons in spin split states may respond faster than majority electrons at non--equilibrium.  The  molecular field acting on the spins delays spin flip transitions $\uparrow \rightarrow \downarrow$ and thus a response of the hot majority electrons and their energy levels. The closing of the exchange splitting in the electron spectrum of ferromagnetic transition and rare--earth metals, ferromagnetic semiconductors, spin split quantum well states in thin ferromagnetic films, etc. will reflect this. The time and spin dependent energy shifts of electrons at non--equilibrium may cause interesting behavior, in particular of magnetic tunnel junctions, spin currents etc.. In ferromagnets the moment reversal lifetime of (local) magnetic moments parallel to the global magnetization is larger than of moments pointing in opposite direction. In antiferromagnets such behavior due to asymmetric spin flip transitions may differ.}

\tableofcontents

\section{Introduction}
Spin dynamics of hot electrons in solids is currently studied intensively \cite{1}. One expects on general grounds that the magnetism in excited non--equilibrium solids, ferromagnets, magnetic semiconductors, tunnel junctions and quantum--dots etc., is controlled by energy and angular momentum conservation. The ultrafast fs--time response occurs due to corresponding fast electronic transitions, strong electron interactions, and fast angular momentum transfer which may also involve the system exciting external electromagnetic field, see Maxwell equations\cite{2}. Clearly, the spin
dynamics of the hot electrons is set by the strength of the molecular fields acting on them and angular momentum transfer and thus varies for different electronic states.

At non--equilibrium a temperature $T_{el}$ is quickly established and acts as a control parameter for the non--equilibrium state. This is also largely the case for strongly itinerant magnetism. In ferromagnets with dominant local moments, Heisenberg like ferromagnet, the temperature $T_{spin}$ referring to the magnetic moment disorder is the control parameter. An interesting case occurs when dynamics involves changes of both magnitude and direction of the magnetic moments.

Note, during ultrafast dynamics (of fs--time scale, 20 to 10 fs or less) no electron temperature etc. is established.

Here, we discuss in particular the time dependent response of hot electrons, of exchange split states, bands, different energy shifts all resulting from spin flip transitions between spin split
states. Note, such transitions may also involve polarized light emission. On general grounds one expects that spin flip transitions of hot minority electrons occur first for energetic reasons more frequently. The spin flip transitions act like a hybridization of the spin split states, for illustration see Figs 1 and 2. Obviously, the transition $\downarrow\rightarrow\uparrow$ causes a shift to larger binding energies. For hot majority electrons a corresponding shift occurs to lower binding \cite{3}.
Note, the asymmetry of $\downarrow\rightarrow\uparrow$ and $\uparrow\rightarrow\downarrow$ spin flip transitions
is also reflected by the lifetimes of hot electrons in ferromagnetic metals\cite{1,4}.

Of course, ultrafast photoemission spectroscopy should generally exhibit such behavior, see in particular recent experiments by Weinelt et al.\cite{3}. Weinelt et al. observed such shifts in Gd \cite{3}.

The electron self--energy $\Sigma_\sigma (\varepsilon, t, F, ...)$, ($\Sigma_\sigma = \Sigma_{\sigma^{'}}\int G_{\sigma^{'}} T_{\sigma \sigma^{'}}$, T is the spin dependent t--scattering matrix, F is the light fluence), of the spin flipping electrons describes the dynamic response,
its real part gives the spin dependent electron energy shifts ($ \Delta \varepsilon_{\sigma} \propto Re \Sigma_{\sigma}$) and its imaginary part the corresponding lifetimes ($\tau^{-1}_{\sigma} \propto Im \Sigma_{\sigma}$). For discussion, note approximately $\Sigma_\sigma \propto \mid V_{\sigma,\sigma^{'}}\mid^2 N N$, where $V$ denotes the spin flipping potential and N the averaged density of states \cite{1,4}.

In view of the spin split density of states in Fe, Co, Ni, for example, one expects at non--equilibrium much smaller asymmetry effects for Co and Ni than for Fe. However, for spin split states in thin films, quantum well states, quantum dots, valence states in Gd (spin split states due to exchange coupling of the 5d--valence electrons to the 4f--electron magnetization), in general for narrow spin split bands such asymmetry regarding spin flip transitions may be large.

Then, for large exchange splitting the minority electrons respond faster upon magnetic disorder. The asymmetry increases for increasing molecular field and disappears for vanishing molecular field and long range magnetic order. Also for large electron temperatures ($T_{el}\rightarrow T_c$) this asymmetry should become smaller and disappear. In case of laser light excitations the asymmetry should become smaller for increasing fluence $F$

Aside from density of states effects the same magnitude is expected for larger times for the energy shifts of minority and majority electrons.
\begin{figure}
\centerline{\includegraphics[width=.5\textwidth]{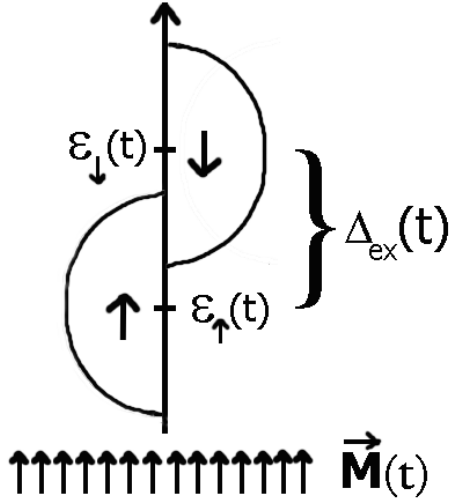}}
\caption{Spin split (states) bands of itinerant electrons. The splitting $\protect\Delta_{ex}(t)=\varepsilon_\downarrow-\varepsilon_\uparrow$ may result from exchange coupling or generally a time dependent molecular field $H_{eff}(t)$ (and possibly including an external magnetic field) acting on the spins of the minority $\protect(\downarrow)$ and majority $\protect(\uparrow)$ electrons. At non--equilibrium spin flip transitions cause a spin dependent shift $\Delta \varepsilon_\sigma$ varying in time t. The magnetization $\protect\overrightarrow{M(t)}$ causes asymmetric behavior of spin flip transitions $\protect\downarrow\rightarrow\uparrow$ and $\protect\uparrow\rightarrow\downarrow$.}
\end{figure}

For times t of the order of the spin lifetime $\tau_\sigma$ of the hot electrons the time dependence of the band shifts $\Delta \varepsilon_\sigma (t)$ should be reflected in electron spectroscopy (of course not for shorter times).

Note in ferromagnetic transition metals like Ni, Fe, Co etc. the lifetimes of the hot electrons resulting for example from laser field excitations are different for minority and majority electrons \cite{1,3,4}. As a consequence different time dependent band shifts $\Delta \varepsilon_\sigma (t)$ and of the center of the exchange split bands occur. The shifts of the electronic states or center of gravity of bands $\varepsilon_\sigma$ are revealed in the time dependence of the exchange splitting $\Delta_{ex}(t) = \varepsilon_\downarrow(t) - \varepsilon_\uparrow(t)$, see Fig.1 for illustration.

Such behavior depends of course on the fluence $F(t)$ of the exciting field and resulting electronic temperature $T_{el}(t)$ or spin disorder temperature $T_{spin}$ in case of local magnetic moments. In case of dominantly itinerant magnetism $\Delta_{ex}$ is proportional to the global magnetization $M(t)$.

Note, the shifts may be calculated using the Hubbard hamiltonian, Fermi--Liquid theory or Green's function methods, see Fig.2,  and the Landau--Lifshitz equation may also be used to calculate $\Delta_{ex}(t) \sim M(t)$.

Regarding spin dynamics and magnetism dynamics at non--equi\- librium
this time dependence is of fundamental interest, since the molecular field felt by the valence electrons varies with electronic states, s, p, d ones etc. \cite{5}. For example the dynamics reflects itinerant vs Heisenberg type magnetism and the behavior of conservation laws for ultrafast responses (reflects the dominant coupling, electron--electron exchange, electron spin--orbit, etc. controlling the angular momentum transfer).

The different shifts of minority and majority electrons result (likely) mostly from the fact that
in the presence of the molecular field spin flip transitions $\downarrow \rightarrow \uparrow$ may occur for energetic reasons
first more frequently than transitions $\uparrow  \rightarrow \downarrow$ for energetic reasons \cite{4}.

\begin{figure}
\centerline{\includegraphics[width=.9\textwidth]{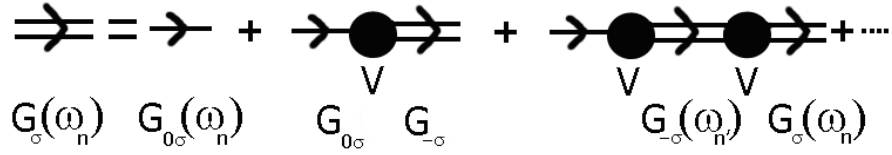}}
\caption{Illustration of Dyson equation for itinerant hot electrons with spin $\sigma$. Spin flip transitions of hot electrons amount physically to spin mixing and quasi hybridization of the exchange split states.
Spin reversal against the molecular field is more difficult and occurs less frequently and delayed in time at non--equilibrium. The delay time $\delta$ may be estimated from $\delta \propto \tau_\uparrow - \tau_\downarrow$, where
$\tau_\sigma$ are the hot electron lifetimes.
This is expected quite generally., for example, in transition metals and rare--earth.
The transition matrix element $V_{\sigma,\sigma^{'}}$ describes transitions between states for spin $\sigma$
and $\sigma^{'}$. In compact form one gets from Dyson eq. the self--energy $\Sigma_\sigma\sim\sum_{\sigma^{'}}\int G_{\sigma^{'}} T_{\sigma \sigma^{'}}$}
\end{figure}
Fig.3 illustrates the asymmetric response (intensity) expected for spin $\downarrow\rightarrow\uparrow$ and
$\uparrow\rightarrow\downarrow$ transitions. Level shifts of minority electrons are expected to occur during times of the order of the demagnetization time
$\tau_M$ ) and hence faster than those of majority electrons.
\begin{figure}
\centerline{\includegraphics[width=.5\textwidth]{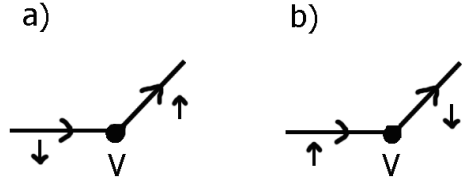}}
\caption{Spin flip transitions of the (hot) excited electrons. The dots refer to the scattering potential V , or more exactly to the t--matrix T, causing spin flip transitions, yielding the spin dependent lifetimes of the hot electrons. Note, the lifetimes of the excited electrons follow from $\tau^{-1}_\sigma \propto \mid V\mid^2$. The molecular field $H_{eff}(t)$ causes generally $\tau_\uparrow  \neq \tau_\downarrow$. For hot electrons
transition a) is favoured relative to transition b). For vanishing molecular field one has $\tau_\uparrow = \tau_\downarrow$.}
\end{figure}

Note, the shifts $\Delta\varepsilon_\sigma$ of the electronic states due to spin flip scattering of the hot electrons are physically quasi spin hybridization, mixing effects, hence $\varepsilon_\uparrow \rightarrow \varepsilon^0_\uparrow +
\alpha ( \varepsilon_\downarrow - \varepsilon_\uparrow )$ (Note, one may attempt to use Kramers--Kronig like analysis to relate such shifts to electron lifetimes). We illustrate the general physical behavior in Fig.4.

\begin{figure}
\centerline{\includegraphics[width=.9\textwidth]{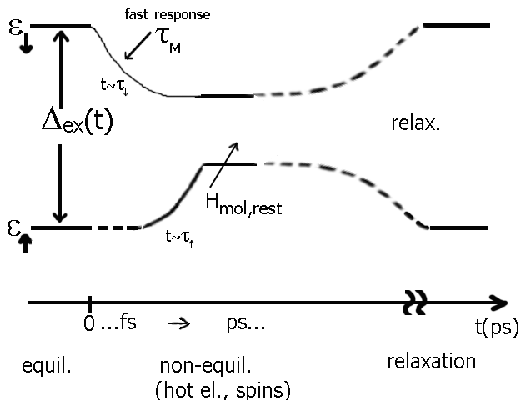}}
\caption{Illustration of time dependence of exchange splitting of magnetic systems at non--equilibrium. Excited minority electrons respond faster than majority ones in ferromagnets. The time delay of a few hundred fs reflects the for energetic reasons different occurrence of spin flip transitions $\downarrow \rightarrow \uparrow$ and $\uparrow \rightarrow \downarrow$. One estimates a spin flip transition energy difference of about $\Delta_{ex}$ or less. The spin dependent response occurs during at most a few hundred fs--times ($t \sim \tau_M \sim \frac{1}{T_c}$) and the delay time is correspondingly shorter than the demagnetization time. The molecular field $H_{mol}(t)$ determines the dynamics and (angular momentum transfer) the time scales. Note, the spin splitting $\Delta_{ex}(t) = \Delta_{ex}(t, H_{mol}(t))$ and its reduction due to hot electrons may saturate if a rest molecular field $H_{mol, rest}$ resulting from magnetization of different electrons of another band or an external magnetic field is present. For example, for Gd the exchange splitting reduces within 1ps from 0.74 eV to 0.6 eV, see experiment by Weinelt et al.. The recovery of the equilibrium magnetization occurs during times controlled by angular momentum transfer and may involve hysteresis due to
magnetic anisotropy. Note, recovery (relaxation) may be relatively slow if angular momentum transfer is slow. For recent experimental results obtained for Gd see Weinelt et al.}
\end{figure}

The demagnetization response as exhibited by the shifts is relatively fast in transition and rare--earth metals, a time scale of a few to hundred fs \cite{6}.
Note, one estimates a demagnetization time $\tau_M \propto (1/T_c)$ \cite{1}. Thus one expects for ferromagnetic Gd with
ordered 4f--electron spins and exchange coupled 5d,6s valence electrons in exchange split states and $T_c = 293K$ a slower demagnetization than for Ni with $T_c = 631K$, Fe etc..

The delay between minority and majority electron response at non--equilibrium is expected to increase with increasing molecular field. For example, the response delay is expected to be larger in Fe than in Ni \cite{4}. Delay times may also result from an external magnetic field.

Of course, the delay depends on the non--equilibrium state, on the density of excited electrons, thus on the fluence of the exciting laser field, $T_{el}(t)$, and excitation energies. The occurrence of majority electron spin flip transitions determines the onset of majority hot electron response. The response of the minority (hot) electrons is expected to occur
during times of the order of the demagnetization time and its dependence on $M(T_{el})$ and thus on $T_{el}$ need be studied and is of interest regarding comparison of theory and experiment \cite{6,7}.

Note, viewing the demagnetization as a fluctuation in magnetic energy ($\Delta E_M$) one is tempted to use for an estimate
of the electronic temperature dependence of the demagnetization time $\tau_M \Delta E_M \sim h$. Possibly this is in accordance with $\tau_M \propto a(T_{el})/T_c$ \cite{1}. Here, the coefficient a depends on density of states changes near $\varepsilon_F$ etc.upon varying $T_{el}$ and needs be calculated carefully using an electronic theory or using for example the Landau--Lifshitz equation. Note, $\frac{dM}{dt}\approx \frac{dM}{dT_{el}}\frac{dT_{el}}{dt}$ and here one may use for the magnetization $M(t_{el})$. The Landau--Lifshitz equation need be extended in general to include both directional and amplitude changes of $\overrightarrow{M(T_{el},..)}$ \cite{8}.  Of course, the dependence of the demagnetization time on fluence sheds also light on its temperature dependence.

Writing for the energies of the valence electrons
\begin{equation}
   \varepsilon_{i\sigma} = \varepsilon^0_i - \sigma J_{eff}M(T_{el},..) + \mbox{spin--flip--scattering--terms},
\end{equation}
suggests that the spin flip transitions cause the asymmetric response of hot minority and majority electrons. This asymmetry is expected to disappear for times larger than the spin dependent lifetimes of the hot electrons (see Fermi--liquid theory and exchange type coupling) and for vanishing molecular field (as $T_{el}\rightarrow T_c$). Thus, as the exchange splitting $\Delta_{ex}(t)$ closes,
the magnitude of the level shifts $\varepsilon_\downarrow$ and $\varepsilon_\uparrow$ become equal.

As suggested by the Hubbard hamiltonian
\begin{equation}
    H = H_0 + \sum_{i,j} U_{ij}n_{i\sigma}n_{j\sigma^{'}}+ ...,
\end{equation}
at strong non--equilibrium with many excited electrons time and spin dependent energy shifts result from level occupation changes $\Delta n_{i\sigma}(t)$ \cite{1}, $\Delta \varepsilon_{i\sigma} \propto \sum_j U\Delta n_{j-\sigma} (t) + ...$.

Physically the spin flips amount to probing of the hot electrons of the spin split states with energy $\varepsilon_{i\uparrow}$ and
$\varepsilon_{i\downarrow}$. Since $\tau_\uparrow > \tau_\downarrow$, first $\Delta \varepsilon_{i\downarrow}(t)$ increases with a rate $\tau_\downarrow \sim \tau_M$, then later during times $t\sim \tau_\uparrow$ one gets that $\Delta \varepsilon_{i\uparrow}(t)$ increases. For times $t>\tau_\uparrow$ one gets $\Delta\varepsilon_{i\uparrow}=\Delta\varepsilon_{i\downarrow}$.

The used physical model suggests that the shifts are proportional to $\Delta_{ex}$ and depend on the lifetimes $\tau_\sigma$ and thus occur first for minority and somewhat later for majority hot electrons. In magnitude both shifts should become equal at times larger than the lifetimes of the hot electrons.

Furthermore, similar behavior and level shifts occur also due to reversal transitions of magnetic moments. Thus level shift dynamics should reveal also itinerant vs. local moment (Heisenberg type) behavior. For example, the energy levels of d--electrons in ferromagnetic transition metals, Ni, Fe,  etc., shift upon reversing the magnetic moment with respect to the (global) magnetization \cite{8}.

Possibly interesting spin dynamics involving level shifts may also occur upon applying a temperature gradient (spatial variation of hot electron density) due to the corresponding spatial variation of the molecular field and for spin currents flowing through a tunnel junction consisting of (two) different ferromagnets, FM$_{1}$ / FM$_{2}$, or a FM / AF tunnel junctions.

Note, the driving force $\Delta \frac{\mu_\sigma}{T_{el}}$ may cause interesting magnetoelectronics effects, see Bennemann
\cite{9}.

In the following we describe briefly the theory for the spin dependent response due to hot electrons. For the analysis one may use for example Keldysh type
Green's functions \cite{10}. Also the Landau--Lifshitz (Gilbert) equation is a general basis for magnetization dynamics \cite{1}. Note, the Landau--Lifshitz equation can be extended to include both time dependent directional and amplitude changes of the magnetization \cite{11}. This may be related then to the electronic theory treating
magnetism dynamics in metals with mixed itinerant and local moment behavior like in Fe and other transition metals or rare earth like Gd \cite{1,8}.

\section{Theory}
We discuss briefly the calculation of the electron spin dynamics and energy shifts at non--equilibrium using various methods like Fermi liquid theory, Green's function theory and using Hubbard hamiltonian taking into account both changes of the direction and amplitude of the magnetization at non--equilibrium.

\subsection{Fermi--Liquid Theory, Hubbard Hamiltonian}

Using Fermi--liquid theory one gets for the system of Fermions at non--equilibrium approximately
\begin{equation}
   \varepsilon_{p,\sigma}(t) =  \varepsilon_{p,\sigma}^0 (0) + tr \int \frac{d^3p^{'}}{(2\pi)^3} f_{p\sigma,p^{'}\sigma^{'}}(t) \delta n_{p^{'}\sigma^{'}}(t) +\ldots \quad .
\end{equation}
Note, the last term of this Eq. can be rewritten using molecular field theory ($f=f_1 + \sigma_i\bullet \sigma_j f_2$ + ...). In general it is for energetic reasons that transitions $\downarrow \rightarrow \uparrow$ occur faster than $\uparrow \rightarrow \downarrow$ ones  (see lifetimes of hot electrons and $\uparrow \rightarrow \downarrow$: $\tau_\uparrow > \tau_\downarrow$, $\downarrow \rightarrow \uparrow$ : $\tau_\downarrow$ shorter) and for the spin flip transition of hot electrons
\begin{equation}
     f(t)_{p\uparrow,p{'}\downarrow}\delta n_{p^{'}\downarrow} \neq f(t)_{p\downarrow,p{'}\uparrow}\delta n_{p^{'}\uparrow},
\end{equation}
as long as the molecular field $H_{eff}$ acting on the electron spins is present, but for $H_{eff}=0$ both scattering amplitudes are equal. Note, $\uparrow$
refers to direction parallel to the magnetization and $\downarrow$ to opposite direction. Spin flip transitions $\downarrow \rightarrow \uparrow$
cause lowering of the energy levels and $\uparrow \rightarrow \downarrow$ cause an increase of the levels, see Fig. 1. The energy shifts are given by 
\begin{equation}
       \Delta \varepsilon_\sigma (t) = \int \frac{d^3p^{'}}{(2\pi)^3} f_{p\sigma,p^{'},-\sigma} \delta n_{p^{'}-\sigma^{'}}(t)+... .
\end{equation}
Here, the dynamics of $\delta n_{p\sigma}(t)$ may be determined using the Boltzmann or Langevin equation \cite{1}. One expects generally in the presence of a molecular field
\begin{equation}
       \Delta \varepsilon_\uparrow (t) \neq \Delta \varepsilon_\downarrow (t)\quad .
\end{equation}
Note, the dependence $f(t,M(t),T_{el},..) \delta n(t, F, ...)$. For times $t \gg \tau_\uparrow,\tau_\downarrow$ one expects $f(t)_{p\uparrow,p{'}\downarrow}\delta n_{p^{'}\downarrow}=f(t)_{p\downarrow,p{'}\uparrow}\delta n_{p^{'}-\sigma^{'}}(t)$  and equal resulting level shifts. First, for times $t\sim \tau_\downarrow\sim \tau_M$ minority electron levels $\varepsilon_\downarrow$ shift and then for times
$t\sim \tau_\uparrow$ levels $\varepsilon_\uparrow$ of majority electrons increase for increasing times. It is $\tau_\downarrow < \tau_\uparrow$ in the presence of a molecular field \cite{4}. The situation is illustrated in Fig.4. Approximately, one may find $\tau_\downarrow \simeq \tau_M$, the demagnetization time. Then the rate of the minority electron level shifts is given by $\tau_M$.

Applying theory also to ferromagnetic Gd with 5d, 5s valence electrons having spin split states due to exchange coupling by the 4f electron magnetization one may compare with experiment by Weinelt et al \cite{3}. Note, experiment observes a fast
minority electron level shift and a delayed (by about 500 fs) response of the majority electron states.

Of course, the delay time $\delta(T_{el}, F,..)$ between level response of minority and majority electrons depends on the molecular field $H_{eff}(t)$, fluence and in general the strength of the magnetic interactions. Approximately, one expects \begin{equation}
\delta \propto (\tau_\uparrow - \tau_\downarrow).
\end{equation}
Hence, the delay time varies in ferromagnetic metals \cite{4}.

Such a delay $\delta(H_{eff}, F,..)$ may also result already in a paramagnet in the presence of a strong external magnetic field.

Note, the spin dependent level shifts
\begin{equation}
    \varepsilon_{p\sigma}(t) -  \varepsilon_{p\sigma}(0)
\end{equation}
are controlled by the fluence F and electronic temperature $T_{el}$ and are expected to be clearly reflected in the time dependence of narrow spin split bands and exchange splitting,
for example in transition and rare--earth metals and ferromagnetic semiconductors. The onset of the minority electron level shifts is expected at $\tau_\downarrow \sim \tau_M (T_{el}$,...) and thus changes in general with $T_{el}$. The response of the majority electrons should occur during times where frequent spin flip transitions $\uparrow \rightarrow  \downarrow$
against the molecular field become possible.

In summary, these shifts are due to the spin flip transitions of the hot electrons. The spin flip transition cause quasi a hybridization of states for spins $\sigma$ and $-\sigma$, see equation.

Using the Hubbard Hamiltonian ($H=H_0 + \sum U n_{i\sigma}n_{i-\sigma} + ...$, including intersite exchange coupling) one
may write $\Delta \varepsilon_{i\sigma} \sim U_{eff}\Delta n_{i-\sigma}$ and then in accordance with Fig.1 at non--equilibrium for the energies ($\varepsilon_{i\downarrow} = \varepsilon^0 + U_{eff}\Delta n_{i\uparrow} +...$)
\begin{equation}
\varepsilon_{i\downarrow} = \varepsilon_{i}^0 + \Delta_{ex}(t)-\Delta \varepsilon_{i\downarrow}(t),  \varepsilon_{i\uparrow} = \varepsilon_{i}^0 + \Delta \varepsilon_{i\uparrow}(t),
\end{equation}
where the shifts occurring at non--equilibrium, from $t=0$ on, result from spin flips of the hot electrons. Note,
$\Delta n_{i\sigma}(t, T_{el})$ could also be calculated using the v.Neumann,  Boltzmann or Langevin equation \cite{12}. It is
\begin{equation}
\Delta \varepsilon_{i\downarrow(\uparrow)} = U_{eff}\Delta n_{i\downarrow(\uparrow)}+...   .
\end{equation}

In view of the mixing of the level for spin $\downarrow$ and $\uparrow$ due to spin flip transitions one expects
\begin{equation}
   \Delta \varepsilon_\downarrow=a(t)\Delta_{ex}(t)+..., \Delta \varepsilon_\uparrow=b(t)\Delta_{ex}(t)+...,
\end{equation}
and $a(t)\neq b(t)$ for $t<\tau_\downarrow, \tau_\uparrow$. For $t>\tau_\downarrow, \tau_\uparrow$ it is $a=b$. Note,
of course the shifts should include all those due to changes in $H_{eff}(t)$ and $M(t)$.

\subsection{Green's Function Theory}
The energy shifts $\Delta \varepsilon_\sigma (t)$ for minority and majority electrons may also be calculated using the Dyson equation for the electron non--equilibrium Green's function $G(t,..)$ or Fourier transform $G(\omega_n)$ \cite{1,4,10}
\begin{equation}
   G(\omega_n)_\sigma = G_{0,\sigma}(\omega_n) + \sum_{n^{'}} G_{0,\sigma}(\omega_n)|V|^2 G_{\sigma^{'}}(\omega_n^{'}) G_\sigma(\omega_n) + \ldots \quad ,
\end{equation}
where $\omega_n = (2n+1)\pi T$ and  V is the matrix element for electron transitions involving spin flips.
This equation is illustrated in
Fig.(2). (Note, Eq. holds also in Wannier representation.)  The energy shifts follow from the real part of the self--energy $\Sigma_\sigma$ and are
given by \cite{4,10}
\begin{equation}
    \Delta \varepsilon_\sigma (t) = Re \sum_{n^{'}} |V|^2 G_{\sigma}(\omega_n^{'}, t, F, T_{el}) + ... \quad .
\end{equation}

Using the Poisson summation formula, see Schrieffer \cite{10}, one gets
\begin{equation}
 \Delta \varepsilon_\sigma (t) = Re \frac{1}{2\pi i k T_{el}} \int_c d\omega^{'}\frac{1}{\exp-(\omega^{'}/kT_{el})+1}
 |V|^2 G_\sigma(\omega^{'})+\ldots \quad .
\end{equation}
Here, we assume that in the non--equilibrium state one may take already, at least approximately, for the temperature
$T=T_{el}(t)$. Also, all shifts due to changes in $H_{eff}(t)$ and $M(t)$ must be included, see shifts due to term
$-\sigma J_{eff}M(t)$.

For further analysis one may use the spectral repr\"{a}sentation of the Green's function and write
$G_\sigma(z)= \int dz^{'} f(z^{'}) A_\sigma (z^{'},...)/(z-z^{'})$ \cite{10}.
For narrow bands with spectral density $A_\sigma(\omega) \sim \delta (\omega- \varepsilon_\sigma)$ one gets then approximately for minority electrons a level shift
\begin{equation}
    \Delta \varepsilon_\downarrow(t) = - a(t,F, T_{el}(t),..) \Delta_{ex}(t)+...,  a \propto |V|^2
\end{equation}
due to scattering $\downarrow \rightarrow  \uparrow$. Similarly one gets for majority electrons due to $\uparrow \rightarrow \downarrow$ transitions
a level shift
\begin{equation}
   \Delta \varepsilon_\uparrow(t) = b(t,F,...) \Delta_{ex}(t,F,..)+..., b \propto |V|^2 \quad .
\end{equation}

In general as indicated by the different lifetimes $\tau_\downarrow$ and $\tau_\uparrow$ for (hot) minority and majority electrons \cite{1,4}, which follow from the imaginary part of the self--energy, $Im \sum_\sigma (\omega)$, one gets
$a(t) \neq b(t)$ and due to the molecular field suppressing if strong enough the transitions $\uparrow \rightarrow  \downarrow$ first a shift of
the minority and then later one for majority electrons for decreasing molecular field. As $\Delta_{ex}(t,T_{el},...) \rightarrow 0$, for vanishing molecular field, one gets $a\rightarrow b$.

\subsection{Theory for Spin--Dynamics involving both directional disorder of the magnetic moments and amplitude changes}

In general one expects that local magnetic moment ferromagnetism (Heisenberg type one) will exhibit a somewhat different dynamical behavior than
itinerant magnetism, since energy scales and thus relaxation times may differ. Note, itinerant magnetism is controlled by intersite electron hopping
and spatially different electron correlations than local magnetic moment magnetism due to strong onsite electronic correlations and intersite exchange coupling.

One needs in general a theory which describes demagnetization at non--equilibrium ( presence of hot electrons ) due to both magnetic moment directional disorder and decrease of the magnitude of the magnetic moments. Thus in general then not only the temperature $T_{el}(t)$ of the hot electrons, but also the temperature $T_{spin}$ referring to the moment disorder controls the dynamics. Hence, one has for the exchange splitting to consider  $\Delta_{ex}=\Delta_{ex}(T_{el},T_{spin})$.

Using the Hubbard hamiltonian
\begin{equation}
       H = \sum_{i,j} t c^{+}_ic_j + \sum_i U \langle n_{i,-\sigma}\rangle n_{i,\sigma} - J\sum_{i,j}\sigma_i\sigma_j +...\quad ,
\end{equation}
which describes electrons hopping between atomic sites i,j and feeling spin dependent effective on--site coupling U, and spin--flip exchange scattering ( J ), etc, one gets for the relative average magnetic moment
\begin{equation}
      \mu(t) = (p^+\mu_+ + p^-|\mu_-|)/ \mu(T=0)\quad .
\end{equation}
Here, + and - refers to magnetic moments pointing parallel and antiparallel to the global magnetization M(t) and p$^{+,-}$ are the probabilities
to find such moments. Furthermore, the relative (global) magnetization is
\begin{equation}
      M (t) = (p^+\mu_+ + p^-\mu_-)/ \mu(T=0) \quad,
\end{equation}
and the long range order parameter is given by
\begin{equation}
     \eta(t)  = p^{+} - p^- \quad .
\end{equation}

Of course, the electron energies change also at non--equilibrium and for sites with moment $\mu_+$ one has
\begin{equation}
      \varepsilon^{+}_\uparrow (t) = \frac{n - \mu_+}{2} U ,\quad    {\varepsilon^+}_\downarrow (t) = \frac{n + {\mu_+}}{2} U .
\end{equation}
Similarly, levels $\varepsilon^{-}_\sigma$ are given, see Moran--Lopez et al., Avignon, Bennemann \cite{8}. The hopping of the electrons in the magnetic moment disordered lattice causes a hybridization of the levels $\varepsilon^{+}_\sigma$ and
$\varepsilon^{-}_\sigma$. The resulting level shifts get more intense as the moment disorder increases.

Note, $U$ could include on--site spin flip effects. The spin splitting is then
\begin{equation}
         \Delta^{+,-} = U \mu_{+,-}+... \quad .
\end{equation}
This is proportional to $M(T_{el},...)$ for itinerant magnetism, but proportional more generally to $M(T_{el},T_{spin},...)$ , if a mixed behavior, both itinerant and local moment one occurs. Note, U is the effective field acting on the moments and plays the role of the molecular field. The
center of gravity of the spin split bands is given by $\varepsilon_\sigma (t) = \frac{1}{W} \int d\varepsilon \varepsilon N_\sigma (\varepsilon)$,
where W denotes the band width and $N_\sigma (\varepsilon)$ the density of states yielding approximately above levels $\varepsilon^{+,-}_\sigma$.

The various properties of the non--equilibrium state may be calculated using Keldysh type non--equilibrium Green's functions \cite{10}. Using the Dyson equation (in tight--binding approximation) one gets
\begin{equation}
     G^i_\sigma  =  G^i_{0,\sigma} + \sum_j G^i_{0,\sigma} t G^j_\sigma +  \sum_j G^i_{0,\sigma} J G^j_{-\sigma} + ...
\end{equation}
where upper index i, j refer to direction of magnetic moment at the corresponding lattice site, t to the hopping integral
and J to an effective spin flip potential. The last term describes the effective hybridization of the spin split states. Lower Wannier type indices referring to lattice sites are not explicitly given and also not the summations over the lattice sites. One may use the Bethe--ansatz and related methods (t--J model etc.)  to determine Green's functions. Note, one may rewrite this Eq. as
\begin{equation}
     G^i_\sigma  =  G^i_{0,\sigma} + \sum_j G^i_{0,\sigma} t G^j_\sigma +
                    \sum_j G^i_{0,\sigma} (J G^j_{-\sigma} \chi J) G^i_{0,\sigma} + \ldots   \quad   .
\end{equation}
In Wannier representation the Green's functions $G^i_{00,\sigma}$, referring to lattice site 0, and $G^i_{01,\sigma}$ to lattice sites 0, 1 etc., may be calculated applying usual techniques and see Avignon, Bennemann, to be published \cite{8}. Above
Dyson eq. extends previous theory of Avignon, Moran--Lopez by including spin--flip transitions.

This theory gives the density of states $N_\sigma(t,\varepsilon)$ and the center of gravity $\varepsilon_\sigma(t)$ of the spin split bands ($\varepsilon_\sigma \sim \int d\varepsilon N_\sigma(\varepsilon)$) \cite{13}.
The free energy at temperature $T_{el}(t)$ is given by $F = E - T S$, where S denotes the entropy approximately given by $S = - k N (p^+ \ln p^+ + p^- \ln p^-)$ \cite{5}.

A detailed study should exhibit differences between the demagnetization of dominantly itinerant ferromagnetism and Heisenberg one, see Avignon, Bennemann \cite{8}.

\section{Results and Discussion}

The main resume of our physical model is that asymmetric response in time of minority and majority spins at non--equilibrium in ferromagnets is  in general expected,
since spin flip processes $\downarrow \rightarrow \uparrow$ and $\uparrow \rightarrow \downarrow$ exhibit different behavior due to the molecular field acting on them. Note, recent experiments in Gd by Weinelt et al. \cite{3} and theory \cite{1,4} suggest such a behavior. Both minority electrons $\varepsilon_\downarrow$ and for magnetic moment disorder reversal of magnetic moments $\mu_{-}$ occurs first in time. Approximately for given temperatures the delay time $\delta$ between minority and majority electron response may be proportional to the molecular field $H_{mol}$. Also for general reasons we estimate $\delta \sim \tau_\uparrow - \tau_\downarrow$ and
$\delta \longrightarrow 0$, as $\Delta_{ex}$ vanishes. Note, presently detailed results for temperature dependence $T_{el}$ of spin flip transitions are not given and would be desired.

One expects that this asymmetric response and time delay of the majority spins is small for Ni, but may play a role already for Fe and other ferromagnets and rare earth, like Gd and others. For increasing laser fluence, $T_{el}$, and density of excitations the delay may get smaller, while larger for increasing molecular field.

Of course, detailed calculations are necessary to determine definitely the asymmetric response in time of hot electrons.
Also, more experiments in various ferromagnets are needed to identify the origin of the fast asymmetric response at non--equilibrium.

In the following we present some preliminary results, see Figs.5 and 7 (see also Avignon and Bennemann \cite{8}).

In Fig.5 we sketch the time dependent behavior expected for exchange split states due to spin--flip scattering in ferromagnetic transition metals and rare--earth.

In response to hot electrons the minority electron states shift first within a time $\tau_{\downarrow}$ and
then later the majority electron states at a time $\tau_\uparrow$. Approximately, it is $\tau_\downarrow \sim \tau_M$. From calculations of the spin dependent lifetimes one estimates
$\tau_\uparrow \approx 2 \tau_\downarrow$ \cite{4}. Hence, one estimates majority electrons respond at times of the order of
$2\tau_\downarrow\approx 2\tau_M$.

For comparison with experiment see recent results by Weinelt et al. for Gd \cite{3}. Then, one estimates minority electrons
\begin{figure}
\centerline{\includegraphics[width=.9\textwidth]{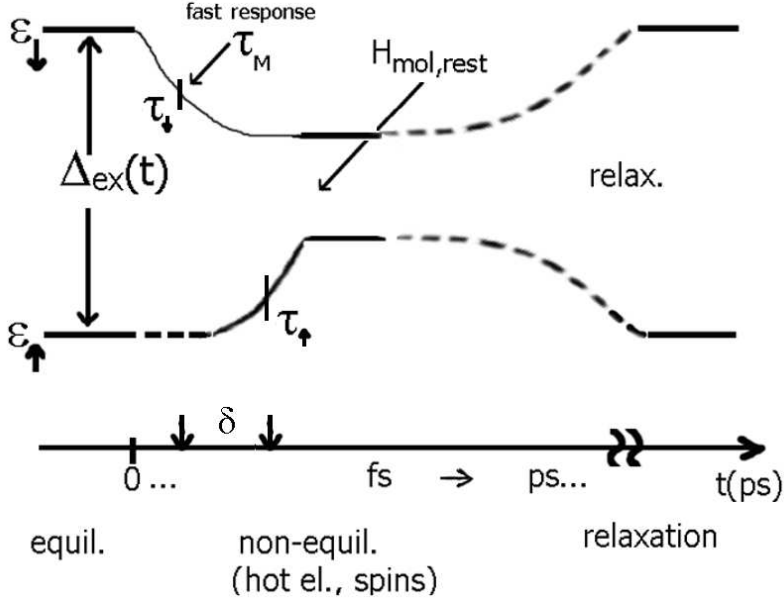}}
\caption{Estimated time dependence of the response of the exchange split states in non--equilibrium ferromagnetic metals with hot electrons. Spin flip transitions cause level shifts $\varepsilon_{\sigma}(t) \propto \Delta_{ex}$. Due to transitions $\uparrow \rightleftarrows \downarrow$ levels $\varepsilon_{\downarrow}$ and $\varepsilon_{\uparrow}$ hybridize, get mixed. For energetic reasons the level $\varepsilon_{\downarrow}$ shifts approximately within the demagnetization time $\tau_{M}$. Then delayed by time
$\delta \approx \tau_{\uparrow}- \tau_{\downarrow}$ the majority electron level $\varepsilon_{\uparrow}$ shifts, approximately also within a time of the order of $\tau_{M}$. The resultant both shifts of $\varepsilon_{\downarrow}$ and
$\varepsilon_{\uparrow}$ are expected to be equal. Note, for ferromagnetic transition metals we estimate $\tau_{M}$
of the order of a few hundred fs. and for Gd etc. of the order of 500 fs.($\tau_{M} \sim \frac{1}{T_c}$). Thus, for Gd
with hot electrons we estimate a decrease of the exchange splitting first due to minority electron level shifts within times of the order of 500 fs and then a decrease due to majority electron level shifts at times of the order of twice $\tau_M$ at about 1 ps. Note, spin relaxation may be slow due to angular momentum transfer.}
\end{figure}
respond at $t\sim 400$ to $500$ fs and majority ones at times of the order of 1 ps.

Note, the delay time $\delta(t,F,..)$ between minority and majority electron response is larger for rare--earth like Gd than for transition metals, $\tau_M \propto \frac{1}{T_c}$ \cite{1}. We assume $\tau_\uparrow \approx 2 \tau_\downarrow$, see calculations of the spin dependent lifetimes of the hot electrons by Zhukov, Knorren et al. \cite{4}.

The decrease of the exchange splitting depends, of course, on the remaining molecular field $H_{mol.}$ after times larger than $\tau_\sigma$. Also onset of minority electron response and delay time $\delta$ depends on light fluence, concentration of hot electrons, electron temperature $T_{el}$.

Note, our estimate for Gd yielding that minority electron levels shift within about 500 fs and majority ones
during about 1 ps is in fair agreement with experiment \cite{3}. In Gd the valence electron states (5d,6s) are spin split
due to exchange coupling $J_{eff}$ by the 4f--electron magnetization. Also $\tau_\uparrow /\tau_\downarrow$ about 1 to 2 is
observed approximately for transition metals \cite{4}.

Using the Hubbard hamiltonian we estimate for the exchange splitting in transition metals
\begin{equation}
     \Delta_{ex} \sim U_{eff}(t, T_{el}) \mu_{av}(t)+\ldots\quad .
\end{equation}
This permits a general test of local moment vs. itinerant magnetism behavior. For example, for Ni one gets after averaging over the directional fluctuations and spatial moment disorder of $\mu(t)$ that
\begin{equation}
U_{eff}(t,T_{el})\mu_{av}(t) \rightarrow 0,
\end{equation}
as $U_{eff} \sim M(t)$ and many hot electrons such that $T_{el}\rightarrow T_c$. Note, in general as $T_{el}$ and molecular field changes both magnitude of the magnetic moments and magnetization vary \cite{8}.

Likely sub--fs spectroscopy will also detect existence of short range magnetic order and of (local) magnetic moments above $T_c$ (or global demagnetization time).
In nearly ferromagnetic metals a time resolved analysis of the magnetic fluctuations might be of interest.

It would be interesting to study also the spin dynamics in antiferromagnets, at surfaces and at interfaces of feromagnetic metals and in alloys and at impurity sites. This sheds more light on how angular momentum transfer controls the spin dynamics.

Fig. 7 shows results for the DOS in a magnetic moment disordered ferromagnet due to mixing of electron states  $\varepsilon_{+,\sigma}$ and $\varepsilon_{-,\sigma}$ at magnetic moment sites (+) and (-). ( Here, (+) refers to magnetic moments pointing into the direction of the magnetization, and (-) to those pointing in opposite direction.)

Note, the electrons move in an "alloy" lattice with lattice sites +, - referring to magnetic moments pointing into direction of magnetization and opposite, see illustration Fig.6 of electrons in a magnetic moment
disordered lattice.
\begin{figure}
\centerline{\includegraphics[width=.3\textwidth]{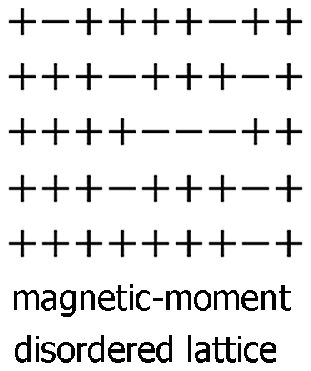}}
\caption{Illustration of magnetic--moment disordered lattice. This amounts to hybridization of the electron states at
atomic sites with magnetic moment pointing into the direction of the magnetization and opposite direction.}
\end{figure}

The above Dyson eq. is used for calculating the electron DOS \cite{8,13}. The probability for a lattice site with +, - depends of course on temperature and magnetization, see Avignon, Moran--Lopez \cite{8}. The magnetic moment disorder causes state shifts like in an alloy amounting to a hybridization of $\varepsilon_{\sigma}^+$ and $\varepsilon_{\sigma}^-$ and which is reflected in the electron DOS at sites + and - . Note, the results do not take into account direct spin--flip transitions due to scattering by the exchange coupling J.

\begin{figure}
\centerline{\includegraphics[width=.5\textwidth]{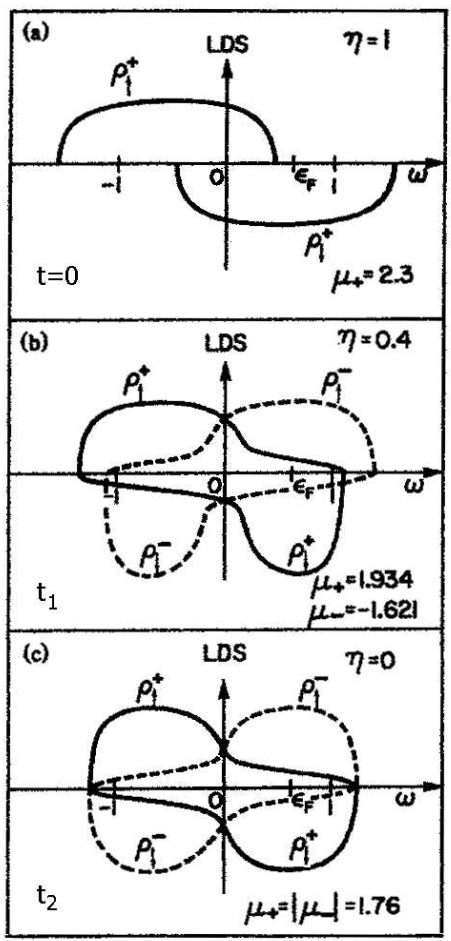}}
\caption{Results for the density of states of majority and minority electrons for different values of the magnetization and magnetic moments. Note, $\rho_\sigma^+$ and $\rho_\sigma^-$ refers to electrons with spin $\sigma$ and to atomic sites with magnetic moment pointing in direction of magnetization (+) and opposite direction (-). $\eta(t)$ refers to the order parameter, magnetization, and $\mu_{+,-}(t)$ to magnetic moments pointing in direction or opposite to the magnetization. Approximately, the time dependence is given by $\eta (t)$ and thus $\eta=1$ corresponds to $t=0$, $t_1$ corresponds to a time of a few hundred fs during which a demagnetization yielding $\eta=0.4$ occurred ($t_1 \sim 0.5 \tau_M$) and $t_2$ with magnetization $\eta=0$ to the demagnetization time. This shows clearly the "alloy" behavior of ferromagnets exhibiting both itinerant and local moment character, mixing of $\varepsilon_{+,\sigma}$ and $\varepsilon_{-,\sigma}$ energy levels.
This implies already similar "alloy" effects regarding mixing of exchange split states due to spin--flip scattering.}                        \end{figure}

Dynamics of the hot electrons and
level shifts due to transitions $\mu_{+} \rightleftarrows  \mu_{-}$ should occur during a few fs or sub--fs. times. Note, the transitions $\mu_{+} \rightleftarrows  \mu_{-}$ occur during a characteristic time expected to be $t < \tau_M$.

Summary:

It is necessary to confirm the physics and approximations used in this discussion by careful
electronic structure calculations
yielding more quantitatively the shifts $\Delta \varepsilon_\sigma (t)$ and the asymmetry of spin flip matrix elements $M_{\downarrow,\uparrow}$ and
$M_{\uparrow, \downarrow}$, see Avignon, Bennemann, to be publ. 2013 \cite{1,4,8}.

Of course spin dependent level shifts in ferromagnets at non--equi\-librium play a role regarding many spin dynamics problems in particular transport ones. A central role for all this in magnetoelectronics \cite{9}, besides electronic theory, plays the Landau--Lifshitz  (Landau--Lifshitz--Gilbert) equation, in particular when damping is important and both directional and amplitude changes occur during non--equilibrium \cite{11,14}, since in general for level shifts
$\Delta\varepsilon_\sigma \propto \frac{dM(t)}{dt}$. A solution in compact form of the Landau--Lifshitz equation was derived by F.Nogueira, K.Bennemann ( to be published FU Berlin, 2013, \cite{14} ).

Non--equilibrium dynamics might offer interesting effects, in particular regarding spin currents in tunnel junctions \cite{15}, photoemission at ferromagnetic surfaces,
spin dependent population dynamics (see already earlier studies by C.Siegmann, W.Eberhard, Bennemann and others). Note,
$n_{\sigma}(\varepsilon,M_\varepsilon(t), t,..)$ might even increase temporarily due to $\tau_\uparrow > \tau_\downarrow$. Note, population dynamics and $n_{\uparrow} (t, \varepsilon,..)\neq
n_{\downarrow}(t,\varepsilon,...)$ in presence of hot electrons will be in general the case. During very short times (likely $t < fs$) it might be possible to observe
induced spin dependent population dynamics of molecular bonds, bond dynamics and of inner core levels of atoms.

Regarding tunnel junctions the spin dependent level shifts, shifts of minority and majority electron bands affect characteristically tunnel currents.
For example, no current flows if on the left side (L) of the tunnel junction the majority band is filled and
the Fermi-level lies in the exchange split minority electron band and on the right side (R) no minority band states are available. Then, in the
presence of hot electrons on the right side of the tunnel junction the resultant shifts of the exchange split bands may permit a minority electron spin
current. For illustration see Fig.8.
\begin{figure}
\centerline{\includegraphics[width=.5\textwidth]{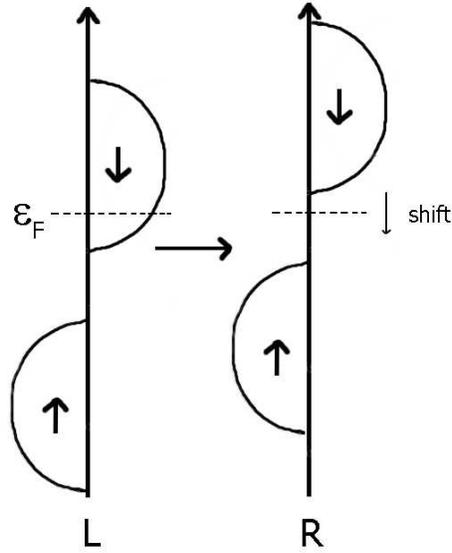}}
\caption{Induced tunnel current (from L towards R) due to band shifts resulting from hot electrons on the right side R of the tunnel junction.}
\end{figure}

The band shifts due to hot electrons may also affect the Josephson like spin currents, see Nogueira, Bennemann \cite{15}. Regarding such currents spin damping, spin lifetimes seem important.

Magnetization reversal will speed up as angular momentum transfer
becomes easily possible, possibly at interfaces, impurity sites and in a.f. or more generally in magnets with inhomogeneous magnetization. This needs more studies.

As discussed using Onsager theory \cite{9} at non--equilibrium magnetoelectronics results for hot electrons from the driving force
$\frac{\Delta \mu_\sigma}{T_{el}}$. Then in particular tunnel junctions may exhibit interesting behavior.

Due to spin, magnetization dynamics electromagnetically induced surface effects occur in topological insulators etc., see Nogueira, Eremin, Bennemann, Meeting DPG, Berlin 2012 and to be publ.\cite{11,14}).

Also, spin dependent behaviour, spin currents, in particular the Josephson like spin current driven by a phase difference,
may be studied using spins of atoms or molecules in an optical lattice. This may help to understand and test in particular various many body theories, the approximations used regarding the Hubbard hamiltonian etc., separation of charge and spin currents.

\section{Acknowledgement}
  I thank C.Bennemann for help in preparing this article.


\begin{thebibliography}{99}
\bibitem{5} One has for the exchange coupling $J(\epsilon_1,\epsilon_2)$, where the $\epsilon_i$ refer to the energies of the interacting electrons. Clearly, strength of exchange coupling may differ for (excited) hot electrons in localized states (d--sates etc.) and for delocalized ones in itinerant s, p-- states.
\bibitem{7} Note, viewing the magnetic energy $\Delta E_M$ as a fluctuation the approximation $\tau_M \Delta E_M \sim h$
      may be used in accordance with $\tau_M \sim T_c^{-1}$. Since $\tau_M^{-1}\sim \int \int d\varepsilon d\varepsilon^{'}|V_{\uparrow \downarrow}|^2 N_\downarrow(\varepsilon^{'})N_\uparrow(\varepsilon)f_\downarrow (\varepsilon^{'})(1-f_\uparrow(\varepsilon))  +...$, see Bennemann \cite{1}, one has $\tau_M=\tau_M (T_{el},..)$.
\bibitem{12} Note, $\Delta \varepsilon_{i\sigma} \sim U_{eff} \Delta n_{-i\sigma} + ...$, and then
      $\Delta \varepsilon_{i\sigma} \sim  -\sigma U_{eff}\Delta M(t,T_{el},..) + U_{eff}\Delta n_{i\sigma}(t)+...$, and here $n_{i\uparrow(\downarrow)}(t)$ may be calculated using the v Neumann, Boltzmann or Langevin equation \cite{10}. $\Delta n_{i\sigma}$ refers to the non--equilibrium state. Obviously, the shifts rates are given by $\tau_\sigma$ and furthermore due to $\tau_\uparrow > \tau_\downarrow$ first minority electron levels shift to larger binding and then majority electron levels shift to lower binding. This states clearly already the asymmetric response in time of minority and majority electrons at non--equilibrium.
\bibitem{13} The DOS $\rho^i_{\sigma} = -\frac{1}{\pi} Im G^i_{00,\sigma}$, $i=+,-$, follows as usual from the Dyson equation for the electron Green's function
      $(\omega - \varepsilon_i^{\sigma}) G^i_{00,\sigma} = 1-zt(p^{(i)i}G^i_{10,\sigma} +
      p^{(i)j}G^j_{10,\sigma}) + ...$.
      Note, missing terms refer to spin--flip scattering. Using Bethe--ansatz one gets similar Eq. for Green's function $G^i_{10,\sigma}=\frac{1}{\omega - \varepsilon_i^{\sigma}}+ ...$, $t$ is the hopping integral, $z$ the coordination no. and $(00)$, $(10)$ are Wannier indices, and conditional probabilities $p^{(i)j}=p^{ij}/p^i$.
\end{thebibliography}
\end{document}